# A second-order self-adjusting steepness based remapping method for arbitrary quadrilateral meshes

Zhiwei He

(corresponding author: he_zhiwei@iapcm.ac.cn)

Institute of Applied Physics and Computational Mathematics, Beijing 100094, China

**Abstract**

In this paper, based on the idea of self-adjusting steepness based schemes[5], a two-dimensional calculation method of steepness parameter is proposed, and thus a two-dimensional self-adjusting steepness based limiter is constructed. With the application of such limiter to the over-intersection based remapping framework, a low dissipation remapping method has been proposed that can be applied to the existing ALE method.

Keyword: ALE, remapping, limiter,

## 1.Introduction

The classical ALE method includes three steps [1]: (1) Lagrange step - In this step, the physical governing equations and the computing mesh are updated, thereby deriving the corresponding Lagrangian step numerical solution and the Lagrangian step computing mesh. (2) Rezone step - In this step, the nodes of the Lagrangian step computing mesh are relocated to a better position to form an optimized computing mesh. (3) Remapping step - In this step, the numerical solution obtained by the Lagrangian step would be shifted from the Lagrangian step computing mesh to the rezoned computing mesh.

This article focuses on the remapping step. Remapping [2,3,4] does not involve physics, and does not include the time evolution of physical quantities. It only maps the physical quantities on the old grid to the new grid, so it is a relatively independent static process. However, the quality of a remapping algorithm directly affects the actual solution of the entire ALE method. Therefore, the study of remapping algorithms has always been a hot but difficult problem.

This article further focuses on how to preserves sharpness (or steepness) of various discontinuous structures during the remapping process, thus to reduce unnecessary numerical dissipation. Recently, the author [5] proposed a class of so-called self-adjusting steepness based schemes. While maintains nominal high order, such schemes can perform anti-diffusion operations on the above discontinuities, thus the resolution of these discontinuities can be improved as much as possible. In this article, the author extends this method to a two-dimensional case. Furthermore, applied to the standard overlay-intersection-based remapping method with linear reconstruction, the author finally obtained a second-order self-adjusting steepness based remapping method suitable for any quadrilateral grid. The final numerical results show that while maintaining the nominal second-order accuracy of the smooth region, the resolution of the discontinuity has been significantly improved.

The framework of this paper is as follows. Section 2 introduces the standard overlay-intersection-based remapping method; Section 3, the author will introduce the newly proposed two-dimensional self-adjusting steepness based remapping method in detail; Section 4, the above new method have been tested using a large number of numerical examples on different types of grids; Section 5 is the conclusion.

## 2. Standard overlay-intersection-based remapping method

Consider two sets of computing meshes: The original (old, Lagrangian) computational mesh whose cell is $c$, volume is $V_c$, and centroid position is $\mathbf{r}_c$; and the new (rezoned) computational mesh whose cell is $\tilde{c}$, volume is $\tilde{V}_{\tilde{c}}$, and centroid position is $\tilde{\mathbf{r}}_{\tilde{c}}$. In this article, it has been assumed that the two sets of meshes have the same convexity, and for mesh cell $c$, its adjacent cell set is defined as $C'(c)$.

Generally, overlay-intersection-based remapping is based on the following formula [3,2]

$$\tilde{c} = \bigcup_{\forall c'} (\tilde{c} \cap c') \qquad (1)$$

This formula provides global remapping between two arbitrary meshes in the same computing zone. If the new mesh is obtained by moving the nodes of the original mesh by a small displacement (such as various mesh smoothing algorithms), we can assume that the intersection operation is just a local operation with cell $c$ and its neighborhood $C'(c)$. Therefore $C(c) = C'(c) \cup c$, and Eq.(1) can be further written as [3,2]:

$$\tilde{c} = \bigcup_{c' \in C(c)} (\tilde{c} \cap c') \tag{2}$$

If two sets of meshes possess the same convexity, the above formula can be written further as follows [3,2]:

$$\tilde{c} = c \cup \bigcup_{c' \in C'(c)} \left( (\tilde{c} \cap c') \setminus (c \cap \tilde{c}') \right) \tag{3}$$

Using this expression, for any scalar quantity per unit volume $f$ ( its mean value on the original mesh cell $c$ is $f_c$ ), its remapping formula can be written as:

$$\tilde{f}_{\tilde{c}} \tilde{V}_{\tilde{c}} = f_c V_c + \sum_{c' \in C'(c)} F_{c,c'} \tag{4}$$

where

$$F_{c,c'} = \int_{\tilde{c} \cap c'} f(\mathbf{r}) dV - \int_{\tilde{c}' \cap c} f(\mathbf{r}) dV \tag{5}$$

In order to perform the above remapping operation, it is necessary to reconstruct the physical distribution $f(\mathbf{r})$ based on the cell average value $f_c$ on the old mesh.

## 3. Second order self-adjusting steepness based remapping method

This paper considers piece-wise linear reconstruction:

$$f(\mathbf{r}) = f_c + \nabla f \cdot (\mathbf{r} - \mathbf{r}_c) \tag{6}$$

However, the above results cannot guarantee the essential non-oscillation property and bound of the final solution. Therefore, the gradient on this cell $c$ needs to be limited.

Under the MUSCL framework [26], the construction of multi-dimensional limiters can generally be divided into two types [27,23]: monoslope method and multislope methods. For simple, efficient and better matching of the intersection algorithm in the remapping method, we adopts the monoslope method, i.e., given a cell $c$, there is only one unique gradient:

$$f(\mathbf{r}) = f_c + (\nabla f)^{\lim} \cdot (\mathbf{r} - \mathbf{r}_c) \tag{7}$$

where $(\nabla f)^{\lim}$ is the final gradient obtained by utilizing various nonlinear limiters. Frequently used limiters of this type are: Barth-Jespersen limiter [15], Venkatakrishnan limiter [16], etc.

However, the above-mentioned commonly used limiters will, in turn, severe smear various discontinuities, especially linear discontinuities (such as contact discontinuities in compressible fluids, diffusive interfaces in multi-material fluids, etc.) And as the calculation time increases, this smearing effect continues.

In the previous references [5], the author proposed a new concept called steepness-adjustable limiters. The biggest character of this class of limiters is that they should have a steepness parameter $\beta$ that provides a mechanism to accurately solve both smooth and discontinuous solution by adjusting $\beta$ according to the flow structure. In that paper, the author propose to a simple method to construct such limiters: extend some existing total-variation-diminishing (TVD) limiters into steepness-adjustable limiters. To capture both smooth and discontinuous solutions, the classic harmonic a limiter [26] is modified into a limiter with adjustable steepness:

$$\Phi(\phi) = \frac{\phi + |\phi|}{1/\beta + \phi} \tag{8}$$

where $\phi$ is the ratio of the upwind increment to the local increment in the standard TVD schemes, and $\beta$ is the steepness parameter. It can be proved theoretically that the steepness-adjustable limiter achieves second-order accuracy when the steepness parameter $\beta$ takes a specific value (remarked as $\beta = \beta_s$). However, when the value of $\beta$ increases (remarked as $\beta = \beta_l$), the slope adjustable limiter introduce the anti-diffusion. In the reference [5], related works have been applied to the flux-splitting based finite difference method, and the conclusions are verified by the numerical test results. Further details on this topic can be referred from the author's previous articles [5].

This article plans to extend such method to the two-dimensional case in order to reduce the numerical dissipation during the remapping process. To constructing such type limiter, the following core issue needs to be addressed: how to give a method to calculate the steepness parameter $\beta$ of a mesh cell (i.e.: there is only one steepness parameter corresponding to a mesh cell). Assuming that the steepness parameter of the cell is known, we can directly propose a multidimensional self-adjusting steepness based limiter. The details are as follows:

Taking the mesh cell $c$ into consideration, the set of edges is denoted as $E(c)$, the total number of edges of this cell is $E_c$. For any edge $e$, its length is $l_e$, and the adjacent cell sharing this edge is $c*$. The smoothness indicator corresponding to this

edge can be defined as:

$$IS_e = \left(f_{c*} - f_c\right)^2 \qquad (9)$$

The linear weight corresponding to this edge is defined as the ratio of the length of this edge to the sum of the lengths of all the sides of this cell, which is represented as:

$$D_e = \frac{l_e}{\sum_{\forall e' \in E(c)} l_{e'}} \qquad (10)$$

Further, the WENO-JS methodology [25] is used to compute the non-linear weight corresponding to this edge:

$$\omega_e = \frac{\alpha_e}{\sum_{\forall e' \in E(c)} \alpha_{e'}} \qquad (11)$$

where

$$\alpha_e = \frac{D_e}{IS_e^p} \qquad (12)$$

where $p$ is the power parameter whose value is usually 2. The final steepness parameter of this cell is defined as:

$$\beta_c = \eta_c \beta_s + (1 - \eta_c) \beta_l \qquad (13)$$

where $\eta_c$ can be defined as:

$$\eta_c = \frac{\prod_{\forall e' \in E(c)} \omega_{e'}}{\prod_{\forall e' \in E(c)} D_{e'}} \qquad (14)$$

Substituting the above formulas (Eqs.9-12) into Eq.(14), the final equation can be derived as:

$$\eta_c = \frac{2^{E_c} \left(\prod_{\forall e \in E(c)} IS_e^p\right)^{E_c - 1} + \varepsilon}{\left(2 \sum_{\forall e \in E(c)} \left(D_e \prod_{s \in E(c) \setminus e} IS_s^p\right)\right)^{E_c} + \varepsilon} \qquad (15)$$

where $\varepsilon$ is a small parameter ,used to avoid the zero value of denominator, usually its value are $10^{-20} \square 10^{-40}$.

Finally, we give the complete process of multi-dimensional second-order self-steepness based remapping method:

(1) In the mesh cell $c$, perform a linear reconstruction to get $\nabla f$, and form

$$f(\mathbf{r}) = f_c + \nabla f \cdot (\mathbf{r} - \mathbf{r}_c).$$

(2) Calculate the steepness parameter of this cell: $\beta_c$.

(3) At each node of this cell $n \in N(c)$, where $N(c)$ is the set of nodes of this cell, calculate an allowable maximum $\phi_n$ as

$$\phi_n = \begin{cases} \dfrac{f_{max} - f_c}{2(f_n - f_c)}, & if\ (f_n - f_c) > 0 \\ \dfrac{f_{min} - f_c}{2(f_n - f_c)}, & if\ (f_n - f_c) < 0 \\ 1, & if\ (f_n - f_c) = 0 \end{cases}$$

where $f_n = f_c + \nabla f \cdot (\mathbf{r}_n - \mathbf{r}_c)$ is the solution of linear reconstruction without limitation at the node $n$. $f_{min}$ and $f_{max}$ are the minimum and maximum values of the current cell and its neighboring cell, respectively.

(4) Set $\Phi_c = \min\limits_{\forall n \in N(c)} (\Phi(\phi_n))$ where $\Phi(\phi_n) = \dfrac{\phi_n + |\phi_n|}{1/\beta_c + \phi_n}$.

(5) Define $(\nabla f)^{lim} = \Phi_c \cdot (\nabla f)$, and the final reconstruction function is

$$f(\mathbf{r}) = f_c + (\nabla f)^{lim} \cdot (\mathbf{r} - \mathbf{r}_c).$$

(6) Execute Eq.(5), and obtain the average value $\tilde{f}_{\tilde{c}}$ on the rezoned mesh cell $\tilde{c}$.

## 4. Numerical examples

In the ALE method, rezone/remapping is not mandatory for every computing step. In principle, the only in the occasion that the Lagrangian computational mesh is kinked, the rezone of the mesh and remapping of the physical quantities are needed. However, rezone/remapped per step (i.e., continuous rezone/remapping) can ensure the convexity of the grid and the intersection operation can be performed locally [3]. Therefore, the continuous rezone/remapping has proved to be a more effective fashion, and is adopted in this article. Additionally, all the computations of this paper are in the following mesh { $x_{i,j}^n$, $y_{i,j}^n$ } are carried on:

$$x_{i,j}^n,\quad i=1,\cdots,i_{max},\ j=1,\cdots,j_{max};\ n=0,\cdots,n_{max};$$

$$y_{i,j}^n,\quad i=1,\cdots,i_{max},\ j=1,\cdots,j_{max};\ n=0,\cdots,n_{max};$$

## 4.1 Tensor Product Mesh

In the computing zone $[0,1]\times[0,1]$, the author uses the following functions [3]:

$$x(\xi,\eta,t)=(1-\theta(t))\xi+\theta(t)\xi^3, \quad y(\xi,\eta,t)=(1-\theta(t))\eta+\theta(t)\eta^2,$$

$$\theta(t)=\frac{\sin(4\pi t)}{2}, \quad 0\le\xi\le 1;\quad 0\le\eta\le 1;\quad 0\le t\le 1.$$

to generate a series of tensor product meshes:

$$x_{i,j}^n = x(\xi_i,\eta_j,t^n); \quad y_{i,j}^n = y(\xi_i,\eta_j,t^n)$$

where $t^n = n/n_{max}, n=0,\cdots,n_{max}$, and

$$\xi_i = \frac{i-1}{i_{max}-1}, \quad i=1,\cdots,i_{max},$$

$$\eta_j = \frac{j-1}{j_{max}-1}, \quad j=1,\cdots,j_{max}.$$

On the basis of these meshes, the author carried out the following three physical quantity distribution tests [3].

### 4.1.1 "Sine" Test

In this example [3], the function $f$ is a smooth function whose expression is:

$$f(x,y)=1+\sin(2\pi x)\sin(2\pi y)$$

where $i_{max} = j_{max} = 65$ and $n_{max} = 320$.

The author uses the Barth-Jespersen limiter and self-adjusting steepness (SAS) based limiter to calculate this example. Note that, in the SAS limiter, the lower bound of the steepness parameter $\beta_l$ is theoretically determined (ensuring second-order accuracy), but its upper bound $\beta_s$ is still uncertain. Therefore, the authors performed the calculation with $\beta_s = 2.9$ and $\beta_s = 4.2$ as an example. Figure 1 give the final results after continuous rezone and remapping, where (a) is the initial field, (b) is the result obtained using the Barth-Jespersen limiter, (c) is the result obtained using SAS limiter ($\beta_s = 2.9$), and (d) is the result obtained using SAS limiter ($\beta_s = 4.2$).

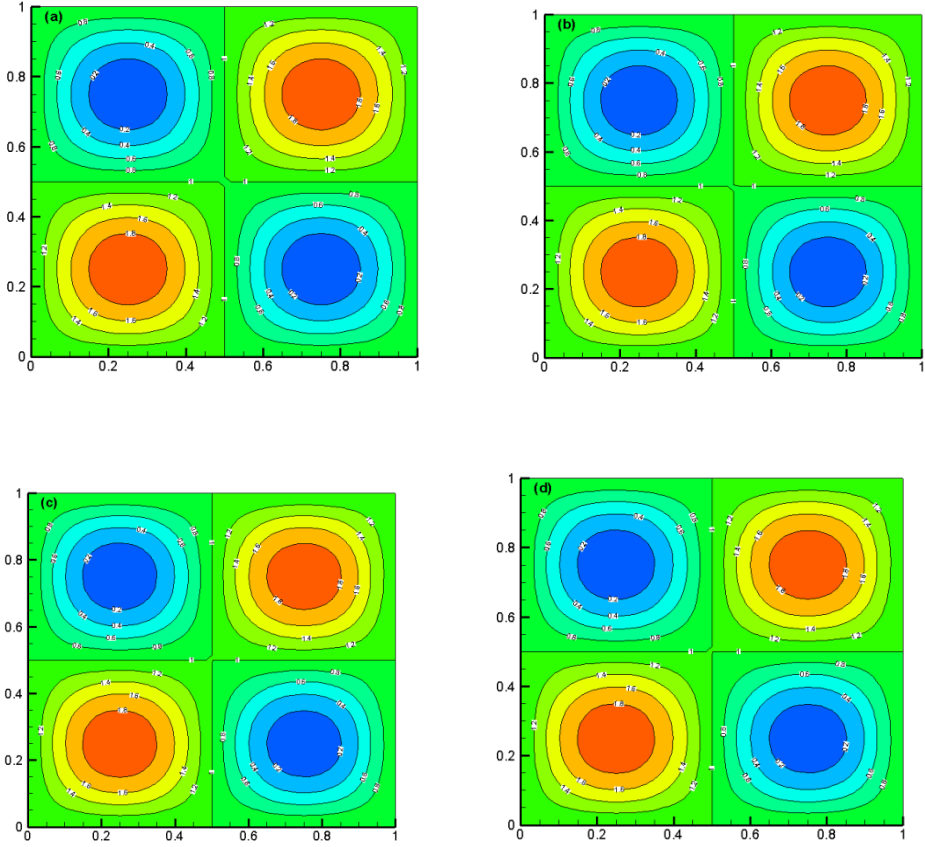

Figure 1. Sine test results on the tensor product mesh. (a) Initial field; (b) Barth-Jespersen limiter; (c) self-adjusting steepness based limiter ($\beta_s = 2.9$); (d) self-adjusting steepness based limiter ($\beta_s = 4.2$).

It can be seen from the figure that no matter what the upper bound of the steepness parameter in SAS limiter is, the final result is very close to the result obtained using the Barth-Jespersen limiter. This result shows that the SAS limiter achieve the nominal second-order accuracy calculation for this smooth structure.

### 4.1.2 "Shock" Test

In this example [3], the function $f$ is a discontinuous function. Its expression is:

$$f(x,y) = \begin{cases} 0, & y > (x-0.4)/0.3 \\ 1, & y \leq (x-0.4)/0.3 \end{cases}$$

where $i_{max} = j_{max} = 65$ and $n_{max} = 320$. Figure 2 give the final results after continuous rezone and remapping, where (a) is the initial field, (b) is the result obtained using the

Barth-Jespersen limiter, (c) is the result obtained using SAS limiter ($\beta_s = 2.9$), and (d) is the result obtained using SAS limiter ($\beta_s = 4.2$).

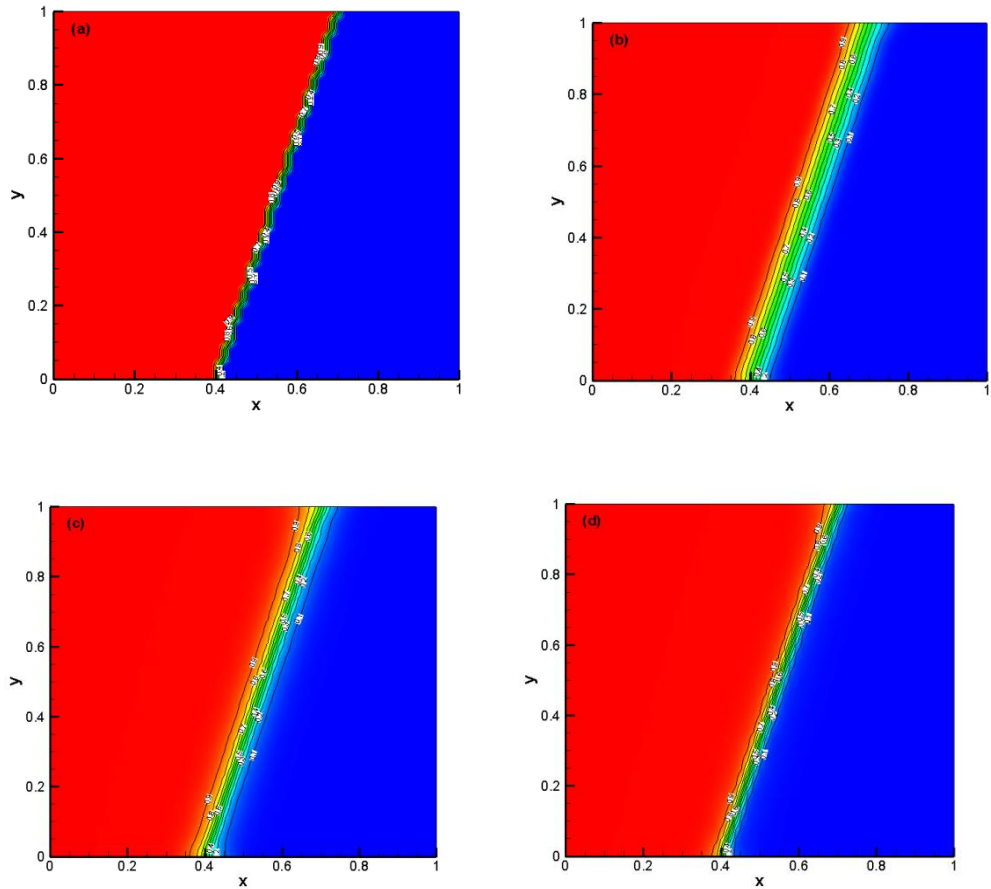

Figure 2. Shock test results on the tensor product mesh. (a) Initial field; (b) Barth-Jespersen limiter; (c) self-adjusting steepness based limiter ($\beta_s = 2.9$); (d) self-adjusting steepness based limiter ($\beta_s = 4.2$).

Comparing the results obtained with the Barth-Jespersen limiter, it can be seen that for discontinuous structures, the SAS limiter has an anti-diffusion mechanism. And as the upper bound of the steepness parameter increases, the anti-diffusion mechanism becomes more obvious, which leads to a significant improvement in the resolution of the discontinuity.

### 4.2 Random Mesh

In this section, the author will consider the type of grid movement that is more in line with the ALE method. On a uniform mesh, the nodes undergo independent

random disturbances to obtain a random mesh [3]:

$$x_{i,j}^n = \xi_i + \gamma \delta_i^n h$$

$$y_{i,j}^n = \eta_i + \gamma \delta_j^n h$$

where $-0.25 \leq \delta_i^n, \delta_j^n \leq 0.25$ is a random number. At the same time, to mimic the rezone process, the author uses the following smoothing method to obtain the new mesh [3]:

$$x_{i,j}^{n+1} = \frac{\left(x_{i-1,j}^n + 2x_{i,j}^n + x_{i+1,j}^n\right) + \left(x_{i,j-1}^n + 2x_{i,j}^n + x_{i,j+1}^n\right)}{8}$$

$$y_{i,j}^{n+1} = \frac{\left(y_{i-1,j}^n + 2y_{i,j}^n + y_{i+1,j}^n\right) + \left(y_{i,j-1}^n + 2y_{i,j}^n + y_{i,j+1}^n\right)}{8}$$

Figure 4 highlights the difference between the initial random mesh and the mesh obtained after different computation steps, where $i_{max} = j_{max} = 17$. In Figure 4, (a) is the initial mesh (black solid line) and the mesh obtained after one step of smoothing (red dotted line); (b) is the initial mesh (black solid line) and the mesh obtained after 20 steps of smoothing (red dotted line).

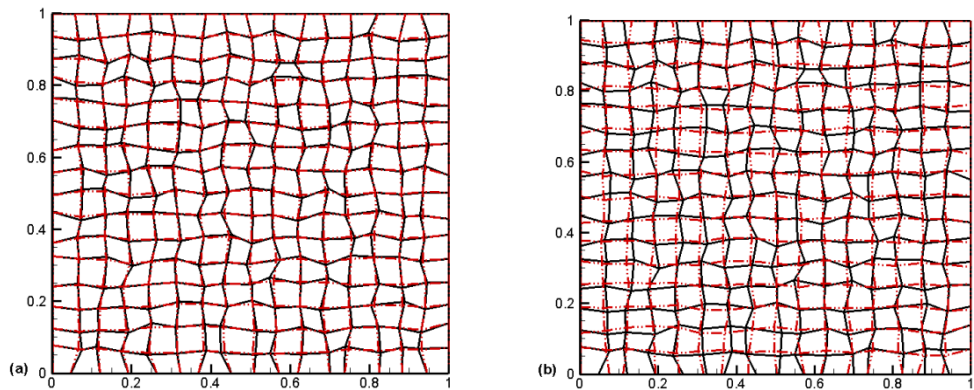

Figure 4. The comparison chart of the difference between the initial random mesh and the mesh obtained after various computation steps, where $i_{max} = j_{max} = 17$. (a) The initial mesh (black solid line) and the mesh obtained after one step of smoothing (red dotted line); (b) the initial mesh (black solid line) and the mesh obtained after 20 steps of smoothing (red dotted line).

On the basis of these meshes, the author carried out the remapping of the above three physical quantity distribution test cases.

### 4.2.1 "Sine" Test

The setup of this example is similar to that of 4.1.1. Figure 3 give the final results after continuous rezone and remapping, where (a) is the initial field, (b) is the result obtained using the Barth-Jespersen limiter, (c) is the result obtained using SAS limiter ($\beta_s = 2.9$), and (d) is the result obtained using SAS limiter ($\beta_s = 4.2$).

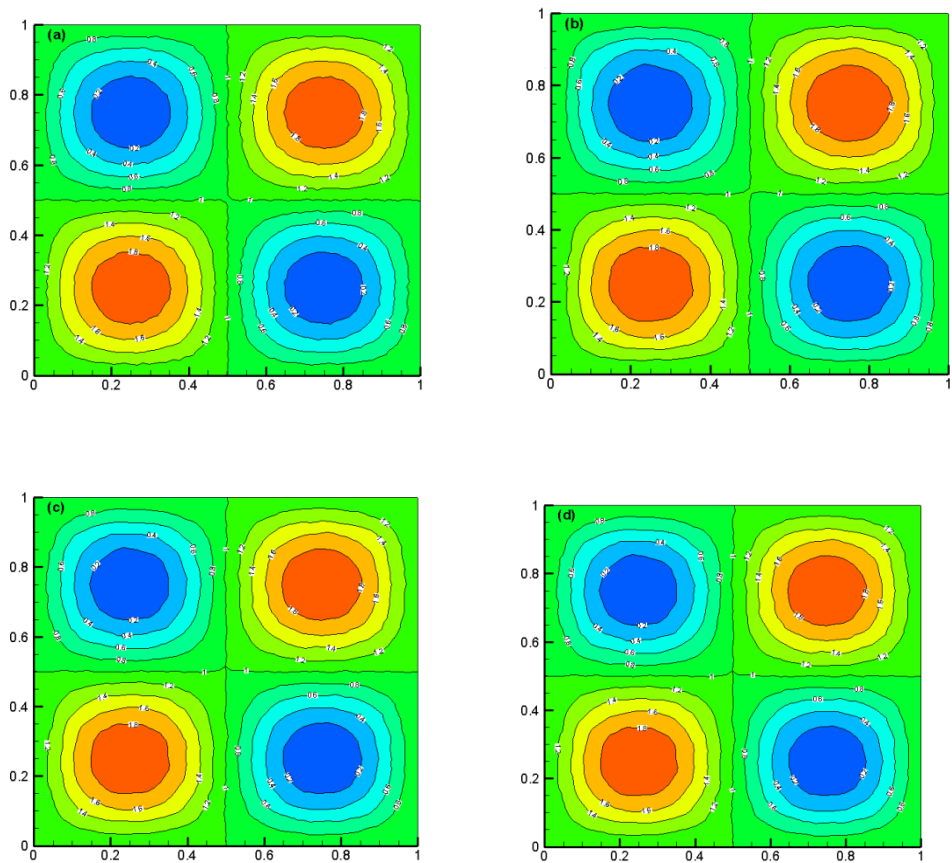

Figure 3. Sine test results on a random mesh. (a) Initial field; (b) Barth-Jespersen limiter; (c) self-adjusting steepness based limiter ($\beta_s = 2.9$); (d) self-adjusting steepness based limiter ($\beta_s = 4.2$).

It can be seen from the figure that even on a random mesh, the final result obtained by using the SAS limiter is still very close to the result obtained using the Barth-Jespersen limiter. This result once again shows that the SAS limiter better guarantees the nominal second-order accuracy for this smooth structure, independent of the type of mesh.

### 4.2.2 Shock Test

The setup of this computing example is similar to that of 4.1.3. Figure 4 give the final results after continuous rezone and remapping, where (a) is the initial field, (b) is the result obtained using the Barth-Jespersen limiter, (c) is the result obtained using SAS limiter ($\beta_s = 2.9$), and (d) is the result obtained using SAS limiter ($\beta_s = 4.2$).

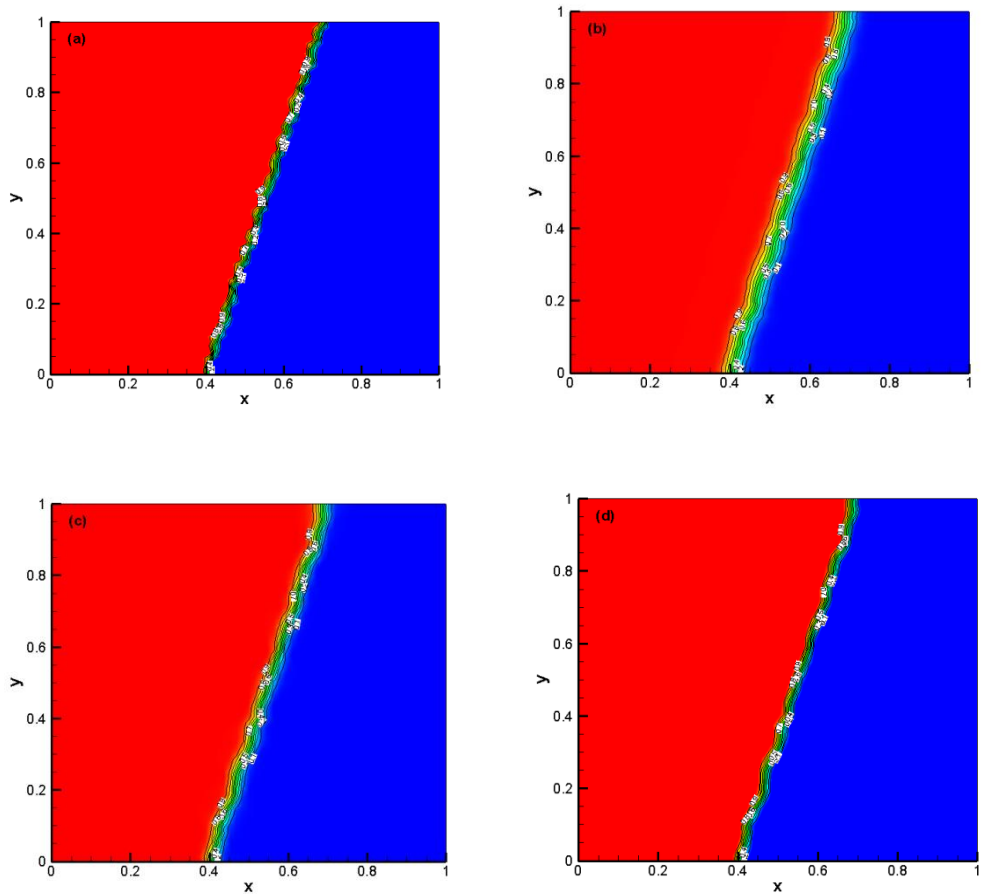

Figure 4. Shock test results on a random mesh. (a) Initial field; (b) Barth-Jespersen limiter; (c) self-adjusting steepness based limiter ($\beta_s = 2.9$); (d) self-adjusting steepness based limiter ($\beta_s = 4.2$).

It can be observed from the figure that with the increase in the upper bound of the steepness parameter in the SAS limiter, the resolution of the discontinuity is significantly improved. This further justifies the above series of conclusions.

### 5. Conclusion

During the remapping process, the physical quantity needs to be reconstructed on the old mesh, and a limiter would be further applied to limit the reconstruction result. However, such methods will bring serious numerical dissipation on some linear discontinuities, such as contact discontinuities in compressible fluids, diffusive interfaces in multi-material fluids, etc. Thus such physical structures are severer smeared. Furthermore, as the calculation time increases, this smearing effect continues.

In the previous references [5], the author proposed a new concept called steepness-adjustable limiters. The biggest character of this class of limiters is that they should have a steepness parameter that provides a mechanism to enable the scheme to accurately solve both smooth and discontinuous solution by adjusting the steepness parameter according to the flow structure. In this paper, we extend such limiter to a two-dimensional case in order to reduce the numerical dissipation during the remapping process. After proposing a method to the steepness parameter of a whole mesh cell, we directly give a multidimensional self-adjusting steepness based limiter. Furthermore, applied to the standard overlay-intersection-based remapping method with linear reconstruction, the author finally obtained a second-order self-adjusting steepness based remapping method suitable for any quadrilateral grid.

The numerical results further suggest that while maintain the nominal second-order accuracy of the smooth region, the resolution of the discontinuity can be significant improved by simply adjusting upper bound of the steepness parameter.